\documentclass[structabstract]{aa}
\usepackage{lscape}
\usepackage{color}
\usepackage{natbib}
\usepackage{graphicx}
\usepackage{natbib}
\begin{document}
\title{\emph{Suzaku} observation of the LINER NGC\,4102}
\author{Gonz\'alez-Mart\'in, O.\inst{1,2}\fnmsep\thanks{\email{omaira@physics.uoc.gr}},
Papadakis, I.\inst{1,2}, 
Braito, V.\inst{3}, \\
Masegosa, J.\inst{4},
M\'arquez, I.\inst{4}, 
Mateos, S.\inst{3},  
Acosta-Pulido, J.A.\inst{5,6},
Mart\'inez, M.A.\inst{4,7}, \\ 
Ebrero, J.\inst{8}, 
Esquej, P.\inst{3},
O'Brien, P.\inst{3},
Tueller, J.\inst{9},
Warwick, R.S.\inst{3}, 
Watson, M.G.\inst{3} }
\institute{$\rm{^1}$ IESL, Foundation for Research and Technology, 711 10, Heraklion,
	      Crete, Greece\\
	      $\rm{^2}$ Physics Department, University of Crete, P.O. Box 2208, Gr-710 03 Heraklion, Crete, Greece\\
	      $\rm{^3}$ Department of Physics and Astronomy, Leicester University, LE1 7RH, UK\\
	      $\rm{^4}$ Instituto de Astrof\'isica de Andaluc\'ia (CSIC), Granada, Spain\\
	      $\rm{^5}$ Instituto de Astrof\'isica de Canarias (IAC), C/Via Lactea, s/n, E-38205 La Laguna, Tenerife, Spain\\
	      $\rm{^6}$ Departamento de Astrof{\'i}sica, Universidad de La Laguna, E-38205 La Laguna, Tenerife, Spain.\\
	      $\rm{^7}$ Grupo de Mec\'anica Espacial and Instituto Universitario de Matem\'atica y Aplicaciones, Universidad de Zaragoza, Zaragoza 50009 Spain.\\
	      $\rm{^8}$ SRON Netherlands Institute for Space Research, Sorbonnelaan 2, 3584 CA,Utrecht, The Netherlands\\
	      $\rm{^9}$ NASA Goddard Space Flight Center, Astrophysics Science Division, Greenbelt, MD 20771. USA\\
}
\authorrunning{O. Gonz\'alez-Mart\'in et al.}
\titlerunning{A \emph{Suzaku} observation of the LINER NGC\,4102}
\abstract
{\emph{Low ionisation nuclear emission-line region} (LINER) nuclei have been claimed to be 
different than other \emph{active galactic nuclei} (AGN) due to the presence of complex absorbing 
structures along the line-of-sight and/or an inefficient mode of accretion onto the supermassive 
black hole. However, this issue is still open.}
{To investigate the broad band X-ray spectrum of NGC\,4102, one of the most luminous LINERs 
in the \emph{Swift}/BAT survey.}
{We studied a 80 ksec \emph{Suzaku} spectrum of NGC\,4102, together with archival \emph{Chandra} 
and \emph{Swift}/BAT observations. We also studied the optical (3.5m/TWIN at Calar Alto observatory) 
and near-infrared (WHT/LIRIS at Observatorio Roque los Muchachos) spectra that were taken 
contemporaneous to the \emph{Suzaku} data.}
{There is strong evidence that NGC\,4102 is a \emph{Compton-thick} AGN, as suggested by the 
\emph{Swift}/BAT detected intrinsic continuum and the presence of a strong narrow, neutral 
FeK$\rm{\alpha}$ emission line. We have also detected ionised Fe$\rm{_{XXV}}$ emission lines in 
the \emph{Suzaku} spectrum of the source. NGC\,4102 shows a variable soft excess found at a 
significantly higher flux state by the time of \emph{Suzaku} observations when compared to 
\emph{Chandra} observations. Finally, a complex structure of absorbers  is  seen with at least two 
absorbers apart from the \emph{Compton-thick} one, derived from the X-ray spectral analysis and 
the optical extinction.}
{All the signatures described in this paper strongly suggest that NGC\,4102 is a \emph{Compton-thick} 
Type-2 AGN from the X-ray point of view. The ``soft excess", the electron scattered continuum component, 
and the ionised iron emission line might arise from \emph{Compton-thin} material photoionised by the 
AGN. From variability and geometrical arguments, this material should be located somewhere between 
0.4 and 2 pc distance from the nuclear source, inside the torus and perpendicular to the disc. The 
bolometric luminosity ($\rm{L_{bol}=1.4\times 10^{43}erg~s^{-1}}$) and accretion rate 
($\rm{\dot{m}_{Edd}=5.4\times10^{-3}}$) are consistent with other low-luminosity AGN. However, the 
optical and near infrared spectra correspond to that of a LINER source. We suggest that the LINER classification 
might be due a different spectral energy distribution according to its steeper spectral index.}
\keywords{galaxies:active - galaxies:nuclei - galaxies:Seyfert - galaxies:individual (NGC~4102) - 
X-ray:galaxies}
   \maketitle
%

\section{Introduction}\label{sec:intro}

Active galactic nuclei (AGN) emit over the entire electromagnetic spectrum and are widely believed 
to be powered by the accretion of matter onto a supermassive black hole 
\citep[SMBH,][]{Rees84}. Several families within the AGN category have been 
established from the observational point of view. Although their classification 
is sometimes misleading, it is widely believed that a unified model can 
explain them all under a single scenario \citep{Antonucci93}. A key ingredient 
in this scheme is a dusty torus whose inclination with respect to the observer's 
line of sight is responsible for the dichotomy between optical Type-1 (with 
broad permitted lines, face-on view) and Type-2 (with narrow permitted lines, 
edge-on view) AGN. However, this scheme needs to be further refined since 
there are several sub-classes of objects that cannot be easily fitted into this 
scenario \citep[for example unobscured Type 2 Seyferts, e.g.][]
{Mateos05,Dewangan05,Panessa02}.

One of the most intriguing cases are \emph{low ionisation nuclear emission-line 
regions} \citep[LINERs, ][]{Heckman80}. As suggested by their low X-ray 
luminosities \citep[$\rm{L(2-10~keV)}$ $\rm{\sim 10^{39-42}erg~s^{-1}}$, see][]{Gonzalez-Martin09A} they could 
be the link between AGN ($\rm{L(2-10~keV)}$ $\rm{\sim 10^{41-45}erg~s^{-1}}$) 
and normal galaxies \citep[$\rm{\sim 10^{38-42}erg~s^{-1}}$,][]{Fabbiano89}. Moreover, they are the dominant 
population of active galaxies in the nearby universe \citep{Ho97}.
However, their nature is not yet well understood. 

Several samples of LINERs have been analysed at X-ray frequencies, a large 
fraction of them showing AGN signatures \citep{Gonzalez-Martin09A,Gonzalez-Martin06,Dudik05}.
In spite of this, it still remains unclear how to 
fit LINERs into the AGN Unification scenario. A radiatively inefficient accretion 
flow onto the SMBH \citep{Ho09a} and/or the presence of highly obscuring matter have been proposed 
to explain the differences between LINERs and more luminous AGN \citep{Goulding09,Dudik09,Gonzalez-Martin09B}.

Using the ratio $\rm{log(Fx(2-10~keV)}$ $\rm{/F([O~III])}$) (R$\rm{_{X/[O~III]}}$, hereinafter),  \citet{Gonzalez-Martin09B}
showed that LINERs have a higher fraction of \emph{Compton-thick} sources than Type-2 Seyfert galaxies.
This implies high column densities and significant suppression of the intrinsic continuum 
emission below 10 keV. Only indirect proof of their Compton-thickness can be obtained with 
\emph{Chandra} and \emph{XMM-Newton} data. Therefore, the 
nature of these sources is yet to be confirmed. A more direct evidence 
comes from the determination of the strength of the neutral iron $\rm{K\alpha}$ emission line
and the direct view of the nuclear continuum above 10 keV. 

NGC\,4102 is a nearby Sb galaxy with a nuclear optical
spectrum that was first classified as an HII region by \citet{Ho97} although its UV 
emission is not compatible with this classification \citep{Kinney93}. \citet{Goncalves99} 
classified its optical spectrum as composite, concluding that the nucleus is
dominated by starburst emission although a weak Type-2 Seyfert component is also present. 
NGC\,4102 is included in the \citet{Carrillo99} sample of LINERs\footnote{This catalogue included all the nuclei
classified as LINERs in the literature to data.} and we have
reclassified it as LINER by means of the emission lines given in \citet{Moustakas06}.

NGC\,4102 has been observed with the {\it Chandra}/ACIS snapshot survey 
\citep{Dudik05}. They classified it as an AGN-like source.
\citet{Tzanavaris07} pointed out its AGN signatures, and considered it as a good candidate for
 harbouring a hidden AGN. They claimed the presence of an iron line, although poor statistics did 
not allow them to accurately constrain its equivalent width. 
\citet{Ghosh08} showed that NGC\,4102
has an AGN and strong star formation activity. They also pointed out the existence
of a reflection component based on a hint of a strong FeK$\rm{\alpha}$ emission line.
According to the $\rm{F_{X}(2-10 keV)/F([O~III])}$ ratio, NGC\,4102
is a good candidate to be a \emph{Compton-thick} source (see Sect. \ref{sec:dis} in this paper).
Therefore, NGC\,4102 is an ideal case to study the obscuration in LINERs.

Here we present the \emph{Suzaku} spectra of NGC\,4102. We also present
optical (TWIN/2.2m in Calar Alto 
observatory) and near infrared (LIRIS/WHT in El Roque de los Muchachos observatory) 
spectra which were taken contemporaneously (up to one month apart) with our \emph{Suzaku} observation.
\emph{Chandra} archival data are also revisited to study the long term variability 
of this source. 

This paper is organised as follows. In Section 2 we describe the X-ray  data reduction and 
observations. In Section 3 we present the X-ray spectral fitting. In Section 4 we 
review the NGC\,4102 activity classification as seen by optical and near-IR observations. 
Finally, we discuss the nature of the emission seen in NGC\,4102 in Section 5 
and report the main conclusions in Section 6. 
A distance of 17 Mpc is assumed for NGC\,4102 throughout the analysis \citep{Tully88}. A $\Lambda$CDM cosmology with 
($\Omega_{\rm M}$,~$\Omega_{\Lambda}$)~=~(0.3,~0.7) and ${H}_{0}$~=~75~ ${\rm km}~{\rm s}^{-1}~{\rm Mpc}^{-1}$
(i.e. z=0.0042) is also assumed.

\section{X-ray observations and data reduction}\label{sec:obs}

\subsection{Suzaku data}

\emph{Suzaku} \citep{Mitsuda07} observed NGC\,4102 for a total exposure time of 
80 ksec on 2009 May 30th. The data were taken using the X-ray Imaging 
Spectrometer (XIS) and the Hard X-ray Detector (HXD) at the HXD nominal 
point\footnote{\emph{Suzaku} data has been obtained centring the source in the nominal position
of HXD in order to maximise the S/N.}. 

For the data reduction and analysis we followed the latest \emph{Suzaku} data
reduction guide\footnote{http://heasarc.gsfc.nasa.gov/docs/suzaku/analysis/abc/}.
We reprocessed all the data files using standard screening within XSELECT 
(``SAA==0" and ``ELV$\rm{>}$5").

The net exposure time of XIS detectors is 79 ksec.
We reprocessed the Spaced-row Charge Injection (CTI) data of the XIS instrument 
using {\sc xispi} task in order to use the latest calibration files at the 
time of writing. We also excluded data with Earth day-time elevation 
angles less than 20$\rm^{o}$ using XSELECT (``DYE\_ELV$\rm{>}$20"). XIS data were 
selected in 3 $\rm{\times}$ 3 and 5 $\rm{\times}$ 5 edit-modes using grades 0, 2, 
3, 4, 6. Hot and flickering pixels were removed using the {\sc sisclean} script. 

Spectra were extracted by using circular regions of 2 arcmin radius\footnote{
This includes $\rm{\sim}$90\% of the emission of the source.} centred in the 
NED nuclear position of NGC\,4102 (R.A. (J2000)= 12:06:23.0 and Dec (J2000) = +52:42:40). 
CXO\,J120627.3+524303 is reported by the \emph{Chandra} Source Catalogue (CSC) as a source within this extraction
region with $\rm{F(0.5-10~keV)=4.8\times 10^{-14}erg~s^{-1}cm^{-2}}$, which
is much smaller than that of NGC\,4102 (see Table \ref{tab:fluxes_components}).

In addition to NGC\,4102, four sources were detected in the XIS field of view: 
CXO\,J120543.3+523806, CXO\,J120548.4+524306, CXO\,J120600.6+523831, and
CXO\,J120633.2+524022. According to the CSC these sources have a 0.5--10 keV
flux of 7.5, 4.8, 11.0, and 6.7 in units of $\rm{10^{-14}erg~s^{-1}cm^{-2}}$. 
Background spectra were extracted using two circles of 2.5 and 1.7 arcmin radii on 
the field, excluding the four sources mentioned above and chip corners to avoid the 
calibration lamps. The response matrix RMF 
and ancillary response ARF files were created using the tasks {\sc xisrmfgen} and 
{\sc xissimarfgen}, respectively. Spectra from the two front illuminated XIS 0 and 
XIS 3 chips were combined to create a single source spectrum ({\sc addascaspec} 
task), while data from the back illuminated XIS 1 chip were kept separate. Both 
resulting spectra were then binned with a minimum of 20 counts in each energy bin in 
to allow the use of $\rm{\chi^2}$ statistics using {\sc grppha} task. 

\emph{Suzaku} HXD/PIN is a non-imaging instrument with a 34' square (FWHM) 
field of view. We reprocessed the HXD/PIN files using standard screening within 
XSELECT (``T\_SAA\_HXD$\rm{>}$500" and ``COR$\rm{>}$8"). We extracted the spectra and 
corrected them for dead-time intervals. We used the variable non-X-ray background 
(NXB) model D \citep[or tuned background, see ][]{Fukazawa09} that the HXD instrument 
team  provides to correct for particle or detector background. This NXB was added 
to the cosmic X-ray background (CXB) to produce the final background spectrum. 
The source was not detected by HXD (see Section \ref{sec:HXD}). 

\subsection{Chandra data}

Level 2 event data from the ACIS instrument were extracted from \emph{Chandra}'s
archive\footnote{http://cda.harvard.edu/chaser/} (ObsID 4014). The source was observed on 2003 
April 3rd as part of a snapshot survey of LINERs by \citet{Dudik05}. The data were 
reduced with the {\sc ciao 3.4}\footnote{http://asc.harvard.edu/ciao} data 
analysis system and the \emph{Chandra}  Calibration Database (caldb 
3.4.0). The observation was processed 
to exclude background flares, using the task 
{\sc lc\_clean.sl}
in source-free sky regions of the same observation. The net exposure time 
after flare removal is 4.9~ksec and the net count rate in the 0.5--10 keV band 
is (6.8$\rm{\pm0.3)\times 10^{-2} counts~s^{-1}}$. 

The source spectrum was extracted from a circular region with 5 arcsec radius centred 
at the same position than the extraction used for \emph{Suzaku}/XIS data. 
The background spectrum was extracted using also a circular region of 18 arcsec 
centred at R.A. (J2000) = 12:06:20.7 and Dec (J2000) = +52:42:13. The net number of 
counts of the spectrum is 340 counts in the 0.5 to 10 keV band. Response and ancillary response files were created 
using the CIAO {\sc mkacisrmf} and {\sc mkwarf} tools. 
The spectrum was binned to give a minimum of 20 counts per bin.

\subsection{Swift/BAT data}

The \emph{Swift}/BAT reduced spectrum was kindly provided to us by the 
\emph{Swift} team. It has been derived from an all-sky mosaic in high energy bins,
averaged over 22 months of data, from 2004 December 15th to 2006 October 27th, 
and extracted from a 17 arcmin circular extraction region \citep[see][for a detailed 
explanation of the data processing]{Winter08,Tueller10}.
NGC\,4102 is reported in the 22 months catalog as one of the 461 sources detected above
a 4.8$\rm{\sigma}$ level in the 14-196 keV band with BAT \citep{Tueller10}. It has been
detected at the 6.96$\rm{\sigma}$ confidence level, and has an average flux of
$\rm{F(14-196~keV)\sim2.2\times 10^{-11} erg~cm^{-2}s^{-1}}$.

\begin{figure}[!t]
\includegraphics[width=0.7\columnwidth,angle=-90]{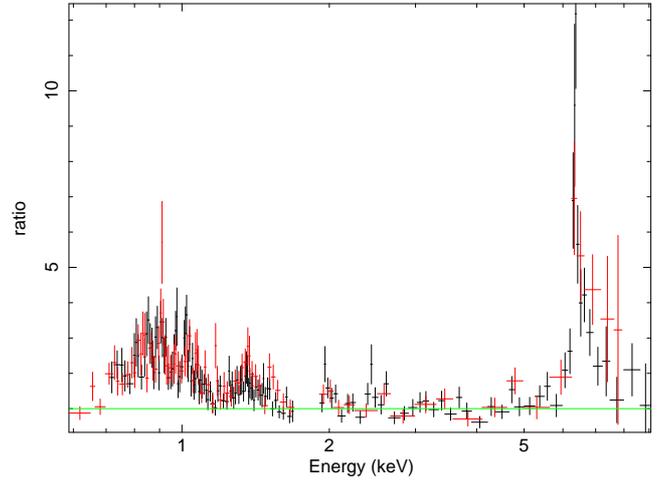}
\caption{Ratio between XIS/\emph{Suzaku} data and a power-law 
model fitted to the 2--6 keV band. Black (dark) points indicate the XIS-0+3 spectrum and red (light)
points the XIS-1 spectrum.}
\label{fig:ratio_pl}
\end{figure}

\section{X-ray spectral fitting of NGC\,4102}\label{sec:spec}

In this section we present the results from various model fits to the 
\emph{Suzaku}, \emph{Chandra}, and \emph{Swift} data. 
All the spectral analysis was done using version 12.5.0 of {\tt XSPEC}. 
All spectral fits include neutral Galactic 
absorption \citep[$\rm{N_{H}(Gal)=1.68\times 10^{20}cm^{-2}}$; ][]{Dickey90}.
Spectral parameter errors are computed at the 90\% confidence level.

\subsection{Suzaku spectra}\label{sec:suzakuspec}

\begin{figure*}[!t]
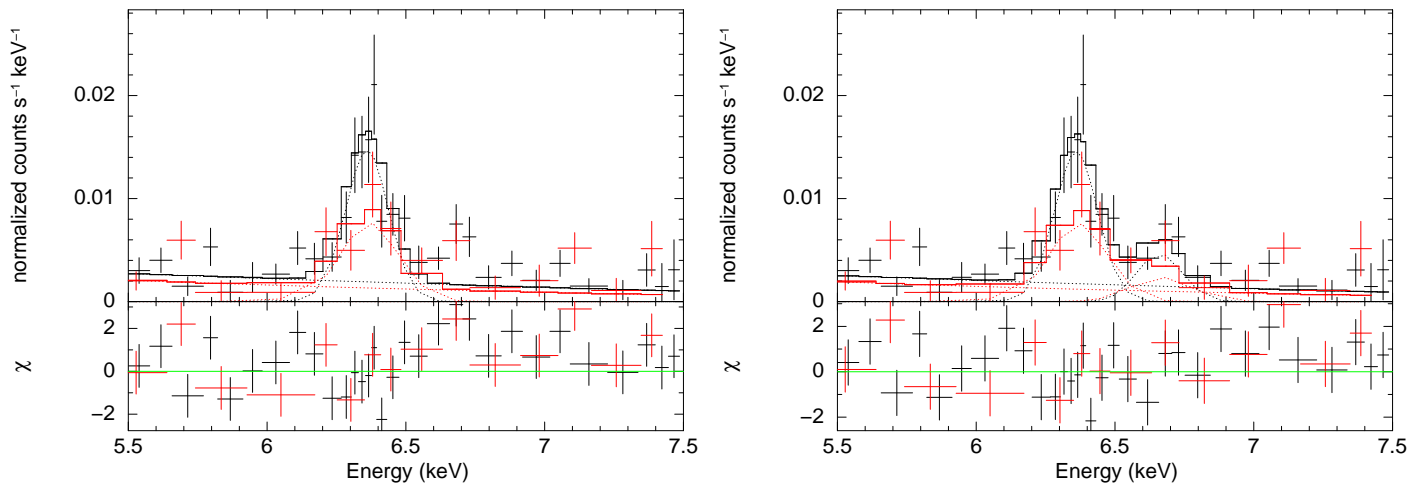

\includegraphics[width=0.7\columnwidth,angle=-90]{N4102_gauss.ps}
\includegraphics[width=0.7\columnwidth,angle=-90]{N4102_2gauss.ps}
\caption{Best-fit (top panels) and residuals (bottom panels) plots of \emph{Suzaku}/XIS data for \emph{Model A} (left) 
and  \emph{Model B} (right) around the Fe emission region
(black and red points as in Fig. \ref{fig:ratio_pl}).
The dotted lines indicate the various model components.
}
\label{fig:Gaussian}
\end{figure*}

\subsubsection{XIS spectra}\label{sec:xisspectra}

\begin{table}[!b]
\begin{center}   
\caption{Results from spectra fitting of the XIS/\emph{Suzaku} data.
}
\begin{tabular}{lccc}
\hline \hline
               			 &		     	 	& 		   		&			        \\
 Model         			 &\emph{A}      		&\emph{B}	   		&  \emph{C}	       		\\
 & \multicolumn{3}{c}{(Models to the 2-9 keV band)}  \\ \hline
$\rm{\Gamma}$  			 & $\rm{1.6_{-0.1}^{+0.1}}$   	& $\rm{1.7_{-0.1}^{+0.1}}$   	&$\rm{2.3_{-0.3}^{+0.3}}$	\\
EW(FeK$\rm{\alpha}$)$\rm{^{(1)}}$& $\rm{1.3_{-0.2}^{+0.2}}$   	& $\rm{0.76_{-0.2}^{+0.2}}$  	&$\rm{0.68_{-0.13}^{+0.11}}$	\\
EW$\rm{(Fe_{XXV})}$$\rm{^{(1)}}$ &  \dots	      		& $\rm{0.18_{-0.07}^{+0.07}}$ 	&$\rm{0.16_{-0.06}^{+0.06}}$	\\
$\rm{\chi^{2}}$/dof 		 & 214.5/164	      		& 188/163 	   		& 181.7/162		       		\\  \hline \hline
               			 &		     	 	& 		   		&			        \\
 Model         			  &  \emph{D}			  &  \emph{E}		         &   \\
& \multicolumn{2}{c}{(Models to the 0.6--9 keV band)} & \\ \hline
$\rm{N_H}$$\rm{^{(2)}}$		  &\dots			  & $\rm{0.7_{-0.5}^{+0.5}}$	 &   \\
$\rm{\Gamma}$  			  &$\rm{2.3^{*}}$		  &$\rm{2.3^{*}}$	   	 &   \\
kT$\rm{^{(3)}}$ 	 	  &$\rm{0.78_{-0.03}^{+0.03}}$    &$\rm{0.73_{-0.03}^{+0.03}}$   &   \\
EW(FeK$\rm{\alpha}$)$\rm{^{(1)}}$ &$\rm{0.68^{*}}$		  &$\rm{0.68^{*}}$	  	 &   \\
EW$\rm{(Fe_{XXV})}$$\rm{^{(1)}}$ &$\rm{0.16^{*}}$		  &$\rm{0.16^{*}}$	  	 &   \\
$\rm{\chi^{2}}$/dof 		  & 456.8/379			  & 445/378		  	 &   \\ \hline
\multicolumn{4}{c}{Models}  \\ \hline
\multicolumn{4}{l}{\emph{A}: \emph{phabs(powerlaw+zgauss)}}	  \\ 
\multicolumn{4}{l}{\emph{B}: \emph{phabs(powerlaw+zgauss+zgauss)}}	    \\ 
\multicolumn{4}{l}{\emph{C}: \emph{phabs({\sc pexrav}+powerlaw+zgauss+zgauss)}}  \\ 
\multicolumn{4}{l}{\emph{D}: \emph{phabs({\sc apec}+{\sc pexrav}+powerlaw+zgauss+zgauss)}}  \\ 
\multicolumn{4}{l}{\emph{E}: \emph{phabs*(zwabs({\sc apec}+powerlaw)+{\sc pexrav}+ zgauss+zgauss)}} \\ 
\hline
\end{tabular}
\label{tab:fittings}
\end{center}
$\rm{^{(1)}}$ EW of emission lines in keV. \\
$\rm{^{(2)}}$ Intrinsic cold absorber column density $\rm{N_H}$ in units of $\rm{10^{21}cm^{-2}}$.\\
$\rm{^{(3)}}$Temperature of the thermal component kT and EWs of the emission lines expressed in keV.\\
$\rm{^{*}}$Fixed parameters to the values obtained in \emph{Model C}. 
See text for details.
\end{table}


NGC\,4102 is detected with the XIS instrument. 
The spectral range of 0.6--9 keV for back-illuminated detectors and 0.7--9 keV for
front-illuminated detectors were used
for the spectral fit of the XIS data. We excluded energies below 0.6 or 0.7 keV 
(for back- or front-illuminated respectively), between 1.7-1.9 keV, and above 9 keV because 
of unsolved calibration uncertainties at these energies and low statistics. 

There are 4600 and 2800 net counts in 
the front- and back-illuminated detectors, respectively. To constrain the shape
of the intrinsic continuum we used the rest-frame 2--6 keV band where we do not expect
contamination from other components usually found in the X-ray spectra of AGN (e.g. soft
excess emission below $\rm{\sim}$2 keV and the FeK$\rm{\alpha}$ emission line at 6.4 keV) and
fitted the data to a power-law model.
The fit is good ($\chi^{2}/dof=129.4/114$), and the best fit 
spectral index is $\rm{\Gamma=1.9\pm0.2}$. 
Fig. \ref{fig:ratio_pl} shows the ratio between the 0.6--9 keV data and this model
(fit statistics of $\chi^{2}/dof=1332/382$). 
The residuals 
plot indicate excess emission below 2 keV and around the 6--7 keV band.

\underline{\it The FeK$\rm{\alpha}$ line}: the excess emission around 6--7 keV 
is consistent with the most 
prominent feature typically observed in the 2--10 keV rest-frame spectra of AGN, i.e. the FeK$\rm{\alpha}$ emission 
line at 6.4 keV. We therefore refitted the 2--9 
keV band data with a powerlaw plus Gaussian line model (\emph{Model A} hereafter). 
The width of the 
Gaussian was fixed to 0.01 keV (i.e. it was assumed to be intrinsically narrow) and the line centroid energy was fixed to 6.4 keV.
Best fit parameters are listed in Tab. \ref{tab:fittings}, and the best fit model
together with the residuals  (in the 5.5--7.5 keV range for plotting purposes), are shown in Fig. \ref{fig:Gaussian} (left).

We also tested the possibility that the line is broad, by letting the width 
of the Gaussian profile vary. The best-fit 
width was $\rm{\sigma(FeK\alpha)=130_{-40}^{+70}}$ eV but the improvement of the fit was 
marginal ($\Delta\chi^{2}=3.7$ for one extra parameter, 
F-statistics probability of 0.09).
 
However, \emph{Model A} is not good enough ($\rm{\chi^{2}_r=1.31}$). 
This is mainly due to the presence of an emission-line feature at $\rm{\sim}$6.7 keV
(see residuals plot in the left panel of Fig.  \ref{fig:Gaussian}). We therefore
included a second Gaussian  (which corresponds to the $\rm{K\alpha}$ line from Fe$\rm{_{XXV}}$) 
to the models with the line energy and width fixed at 6.7 keV and 0.01 keV, 
respectively (\emph{Model B} hereafter). The fit is significantly improved ($\Delta\chi^{2}=26$, 
for one extra parameter when compared to \emph{Model A}). 
The best fit model is shown in Fig. \ref{fig:Gaussian} (right panel) and the best-fit
parameter values are listed in Tab. \ref{tab:fittings}. 
The power-law index is consistent with the previous value. 
We also tested the possibility of this line to be broad by letting
its width to be a free parameter. We get a considerable improvement
on the fit ($\Delta\chi^{2}=13.5$ for one extra parameter,
F-statistics probability of $\rm{6.5\times 10^{-4}}$) with $\rm{\sigma(Fe_{XXV})=390_{-105}^{+450}}$ eV.
The neutral FeK$\rm{\alpha}$ line in this case is still consistent with being narrow [$\rm{\sigma(FeK\alpha)<70}$ eV].

\begin{figure*}[!t]
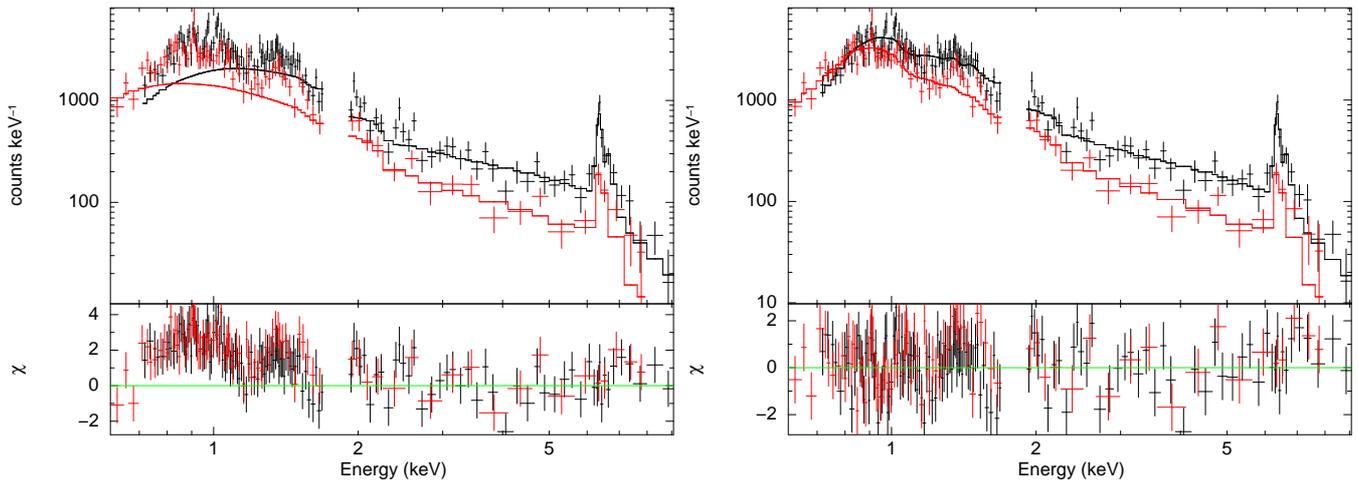

\includegraphics[width=0.7\columnwidth,angle=-90]{NGC4102_modelC_extrapol.ps}
\includegraphics[width=0.7\columnwidth,angle=-90]{NGC4102_modelE.ps}
\caption{(Left): Extrapolation of the 2--9 keV XIS/\emph{Suzaku} best fit model (\emph{Model C}) 
to lower energies. (Right): Final fit (top panel) and residuals (bottom 
panel) to the XIS/\emph{Suzaku} data using \emph{Model E}. See Tab. \ref{tab:fittings} and text for an
explanation on the components used in each model.}
\label{fig:bestfit}
\end{figure*}

\underline{\it The continuum shape in the 2--9 keV band}:
even after the inclusion of the 6.7 keV lines the EW of the 6.4 keV line is 
$\rm{\sim}$680 eV, which strongly suggests a \emph{Compton-thick} source 
\citep[see][]{Maiolino98}. In this case,
we expect the presence of a strong reflection component as well. For that reason, we 
refitted the 2--9 keV spectrum with a model consisting on the two narrow lines at 6.4 keV and 6.7 
keV, a power-law, and {\sc pexrav} in {\tt XSPEC}, which describes reflection from neutral material
\citep{Magdziarz95}. The 
inclination angle of the reflector was set to 60$\rm{^{o}}$ and the iron abundance 
was fixed to 1. {\sc pexrav} was used in such a way that it 
produces the reflected photons only to test if the spectrum is consistent with a 
`pure' reflection model. 
The fit is now acceptable 
(\emph{Model C} hereafter, see Tab. \ref{tab:fittings}) and the best fit spectral index is 
$\rm{\Gamma=2.3\pm0.3}$,  similar to what is observed in other LINERs \citep{Gonzalez-Martin09A}. 
Finally, we allowed to vary the centroids of the Gaussians. The results do not change, showing centroids at E($\rm{FeK\alpha}$)=  $\rm{6.40\pm0.02}$ keV
and E($\rm{Fe_{XXV}}$)$= \rm{6.73\pm0.04}$ keV, consistent with the theoretical energies
of these transitions. We also tested the width of the lines.
The width of the 6.4 keV line is still consistent with zero [$\rm{\sigma(FeK\alpha)<60}$ eV]
while the 6.7 keV line is now also consistent with zero $\rm{\sigma(Fe_{XXV})<860}$ eV. 

\underline{\it The full energy band spectrum}:
our best-fit \emph{Model C} fails 
to describe well the full band \emph{Suzaku} data (see Fig. \ref{fig:bestfit}, left). The spectra show an 
excess below 2 keV.  Many LINERs show a soft excess that 
can be fitted with a thermal model \citep{Gonzalez-Martin09A, Gonzalez-Martin06}. 
Thus, we added a thermal component ({\sc apec} on  {\tt XSPEC}) to the model (\emph{Model D} hereafter). Abundances 
were fixed to the solar value. The spectral index, the parameter values of the reflection
component and of the Gaussian lines were kept
fixed to those in \emph{Model C}.  Best-fit results are listed in Tab. \ref{tab:fittings}. However, the fit overestimates the spectra below 0.7 keV. 
In order to account for this, an additional absorber ({\sc zwabs} in  {\tt XSPEC}) was 
included, absorbing power-law and thermal components. This model properly 
describes the XIS/\emph{Suzaku} dataset 
(\emph{Model E} hereafter, see Tab. \ref{tab:fittings} and Fig. \ref{fig:bestfit}, right). 
The best-fit results and the quality of the fit do not differ 
if we let free the normalization of the Gaussians and the spectral index: 
EW($\rm{FeK\alpha)=700\pm130}$ eV, 
EW($\rm{Fe_{XXV})=170\pm60}$ eV, and $\rm{\Gamma=2.4\pm0.4}$.
The resulting temperature and hydrogen column density $\rm{N_H}$ are consistent with 
those in other LINERs  \citep{Gonzalez-Martin09A}. Note that some residuals still appear
in the 1--2 keV band (Fig. \ref{fig:bestfit}, right bottom panel).
See Section \ref{sec:soft} for a detailed discussion of the soft-excess.

The observed fluxes and absorption corrected luminosities using \emph{Model E} are 
reported in Tab. \ref{tab:fluxes_components}. Sixty per cent of the
0.5--2.0 keV flux is contributed by the power-law, 38\% by the thermal component, and 
2\% by the reflection component. In the hard band (2--10 keV), 54\% of the flux
is contributed by the power-law component, while the reflection component, the 
emission lines and the thermal component contribute 27\%, 17\%, and 2\%, respectively.
The flux of the FeK$\rm{\alpha}$ line using the \emph{Model E} best-fit results is
F(FeK$\rm{\alpha}$)=$\rm{1.5_{-0.2}^{+0.3}\times10^{-13}erg~s^{-1}cm^{-2}}$.

\subsubsection{HXD/PIN spectrum}\label{sec:HXD}

\begin{figure*}[!t]
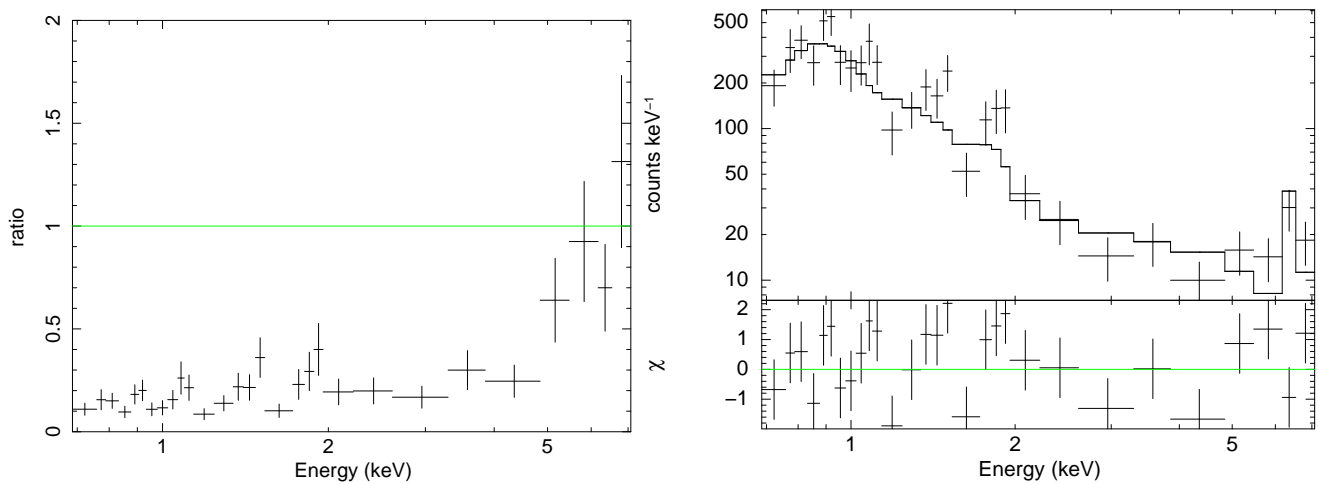

\includegraphics[width=0.7\columnwidth,angle=-90]{N4102_suzakumodel_chandradata.ps}
\includegraphics[width=0.7\columnwidth,angle=-90]{N4102_chandra_factor.ps}
\caption{(Left): Ratio between the ACIS/\emph{Chandra} and the XIS/\emph{Suzaku} data best-fit \emph{Model E}. 
(Right): \emph{Chandra} best-fit (top panel) and residuals (bottom panel) using \emph{Model E} and 
varying the normalisation of the power-law and the thermal components by the same amount. }
\label{fig:ratio_suz_chan}
\end{figure*}

NGC\,4102 is not detected by the HXD/PIN above the total background. The count rate of 
the source is (4.6$\rm{\pm2.5)~10^{-3}}$ counts/sec, only 1\% above the background. 
We have estimated an upper limit for the source flux of F(14-70~keV)$\sf{<}$1.4
$\rm{\times 10^{-11} erg~s^{-1} cm^{-2}}$, from the accuracy of the NXB model. The expected flux 
extrapolating \emph{Model E} is F(14-70~keV)=$\rm{(1.6\pm0.3)\times 10^{-12} erg~s^{-1} cm^{-2}}$, 
consistent with this limit.  

\begin{table}[!b]
\begin{center}   
\caption{Observed fluxes and intrinsic luminosities during the 
	\emph{Suzaku} and \emph{Chandra} observations.
}
\begin{tabular}{lcccccccc}
\hline \hline
Band                              & \multicolumn{2}{c}{Flux}   & & \multicolumn{2}{c}{Luminosity}      \\ \cline{2-3} \cline{5-6} 
   (keV)                           & \emph{Suzaku}  & \emph{Chandra} & &  \emph{Suzaku}  & \emph{Chandra}  \\ 
\hline
0.5--2                    & $\rm{11.4\pm0.5}$       &   $\rm{1.6\pm 0.2}$      & & $\rm{10.7\pm 0.9}$ & $\rm{1.4\pm 0.2}$ \\
2--10                     & $\rm{13.0\pm 0.5}$      &   $\rm{6.1\pm1.3}$       & & $\rm{8.8\pm 0.8}$    &  $\rm{4.1\pm 0.4}$ \\ 
0.5--6                    & $\rm{18.1\pm0.7}$       &   $\rm{3.6\pm0.4}$       & &  $\rm{14.5\pm 0.6}$   & $\rm{2.9\pm 0.3}$ \\
6--9                       & $\rm{5.0\pm0.3}$         &  $\rm{4.0\pm1.5}$        & & $\rm{3.4\pm 0.2}$  & $\rm{2.1\pm 0.5}$ \\ \hline
\end{tabular}
\label{tab:fluxes_components}
\end{center}
Observed fluxes expressed in units of $\rm{10^{-13}erg~s^{-1} cm^{-2}}$ and intrinsic luminosities 
in units of $\rm{10^{40}erg~s^{-1}}$. 
\end{table}

\subsection{Chandra spectrum}\label{sec:chandraspectrum}

Since the \emph{Chandra} spectrum is part of a snapshot survey, its quality 
is poorer than the \emph{Suzaku}/XIS spectra. 
Thus, we decided to fit the \emph{Chandra} spectrum using \emph{Model E} as a baseline.
Fig. \ref{fig:ratio_suz_chan} (left) shows the ratio (\emph{Chandra} data)/(\emph{Suzaku} best-fit \emph{Model E}). 
The source flux was significantly smaller during the \emph{Chandra}
observation.

We use different extraction regions for \emph{Chandra}/ACIS and \emph{Suzaku}/XIS
(5 arcsec and 2 arcmin, respectively). Thus, this differences could be due to 
aperture effects. 
To test this possibility we re-extracted the \emph{Chandra}/ACIS
spectrum using the same extraction region used in \emph{Suzaku}/XIS data. 
The final spectrum is of poorer quality, as expected, so for the spectral fit we used the same model
described for the \emph{Chandra} small aperture data (see below) to obtain flux estimates: 
$\rm{F(0.5-2~keV)=3.3\pm0.3\times10^{-13}erg~cm^{-2}~s^{-1}}$ and
$\rm{F(2-10~keV)=7.1\pm0.7\times10^{-13}erg~cm^{-2}~s^{-1}}$. These
fluxes are still well below the \emph{Suzaku} fluxes. Thus, it is unlikely that
this variation is due to aperture effects.

This flux deficit could be caused either by an increase in the absorption, or a decrease
in the continuum flux. We tested the 
former by fitting \emph{Model E} to the \emph{Chandra}
spectrum, keeping all  the model parameters fixed to the values listed in Tab. \ref{tab:fittings},
apart from $\rm{N_H}$. The best-fit model was unacceptable ($\rm{\chi^{2}/dof=221.9/28}$).
We also tested the addition of cold absorption only to the power-law component
but it cannot reproduce the variations ($\rm{\chi^{2}/dof=1483/28}$) since the flux of the thermal component
obtained by \emph{Suzaku} data is well above the \emph{Chandra} 0.5-2.0 keV
flux. Thus, cold absorption changes alone do not explain the observed variations. 
To test the flux variations we let the normalisations free for the power-law and/or 
thermal components. Changes of either the power-law component normalisation 
or the thermal component normalisation alone could not explain the observed 
variability ($\rm{\chi^{2}/dof=1487/28}$ and $\rm{\chi^{2}/dof=2942/28}$, respectively). 

\emph{Model E} fits well the \emph{Chandra} spectrum if we let free the normalisation of both
components  ($\rm{\chi^{2}/dof=40.7/27}$).
The \emph{Chandra} normalisation of the power-law and thermal components are 
$\rm{12_{-2}^{+3}}$\% and $\rm{13_{-3}^{+6}}$\% of the 
\emph{Suzaku} values, respectively. The variations are consistent 
with being the same for both components. In fact, a satisfactory fit is 
also obtained if we force the same factor of variations in both components
($\rm{\chi^{2}/dof=40.7/28}$). 
In this case, the \emph{Chandra} data require both a power-law 
and thermal component normalisations which are $\rm{12_{-1}^{+2}}$\% of 
that obtained using \emph{Suzaku} data. This best-fit 
model is shown in Fig. \ref{fig:ratio_suz_chan} (right). 

Changes on the spectral index and/or other parameter(s) 
cannot be tested due to the low statistics of the current \emph{Chandra} data set. 
Nevertheless, we tested the strength of the FeK$\rm{\alpha}$ emission line by letting
the normalisation of the Gaussian component free. 
The flux of the FeK$\rm{\alpha}$ line is
F(FeK$\rm{\alpha}$)=$\rm{1.1_{-0.4}^{+0.8}\times10^{-13}erg~s^{-1}cm^{-2}}$ 
($\rm{\chi^{2}/dof=39.4/27}$, see \ref{sec:xisspectra}), consistent with
the flux of the line inferred from the \emph{Suzaku}/XIS data.

Observed fluxes and intrinsic luminosities for the 0.5--2.0 keV, 2--10 keV, 0.5--6 keV, and 
6--9 keV bands are reported in Tab. \ref{tab:fluxes_components}. While the 6-9 keV band flux 
is consistent with \emph{Suzaku} observations, the 0.5--6 keV flux is 5 times lower during 
\emph{Chandra} observation.

\subsection{Swift spectrum}\label{sec:swiftspec}

The extrapolation of the best-fit \emph{Model E} to higher energies predicts a 
flux in the 14--70 keV band of F(14-70~keV)$\sf{\sim 2.1\times 10^{-12} erg~s^{-1} cm^{-2}}$.
This is roughly 10\% of the \emph{Swift}/BAT flux, according to the 22 months catalogue 
of sources \citep{Tueller10}. Thus, a reflection component alone cannot explain
the observed \emph{Swift}/BAT spectrum. 

Since the aperture of \emph{Suzaku}/XIS is smaller than 
that of \emph{Swift}/BAT (2 and 17 arcmin, respectively), this discrepancy could
be due to the difference on the apertures. 
In addition to NGC\,4102, four sources in the XIS field of view
may also contribute to the total \emph{Swift}/BAT flux (namely 
CXO\,J120543.3+523806, CXO\,J120548.4+524306, CXO\,J120600.6+523831, and
CXO\,J120633.2+524022). According to the \emph{Chandra} Source Catalogue (CSC) 
these sources have 0.5--10 keV band fluxes of 7.5, 4.8, 11.0, and 6.7$\rm{\times 10^{-14}erg~s^{-1}cm^{-2}}$. 
Their total flux, $\rm{2.9\times 10^{-13}erg~s^{-1}cm^{-2}}$,
represents a contribution of 10\% to the flux derived for NGC\,4102. 
We extracted the \emph{Suzaku}/XIS spectra of these 
sources following the same procedure as for NGC\,4102. We did not find 
any indication of reflection dominated spectra (no hints of the FeK$\rm{\alpha}$ emission line
and the best-fit spectral index is $\rm{\Gamma \sim 1.6-2.0}$).
In any case, even if we assume the worst case scenario in which all the sources contribute above 10 keV 
with a spectra like NGC\,4102, the expected flux cannot explain the 
excess observed by \emph{Swift}. 

\begin{figure}[!b]
\includegraphics[width=0.7\columnwidth,angle=-90]{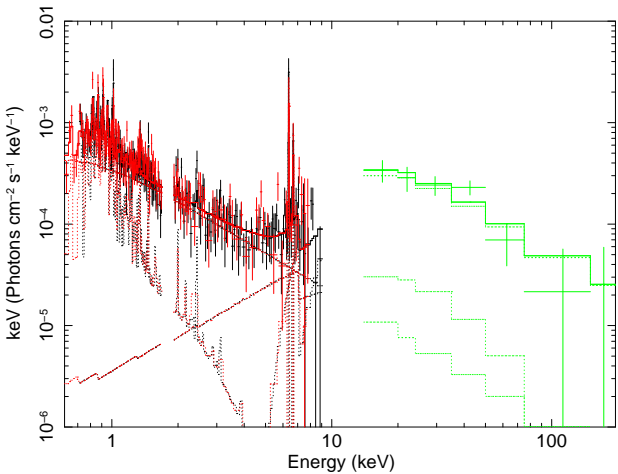}
\caption{Best fit to the \emph{Suzaku}/XIS (red and black) and \emph{Swift}/BAT (green) spectra.}
\label{fig:suz_bat}
\end{figure}

We therefore added a power-law component and absorption to \emph{Model E}, and 
refitted the \emph{Swift}/BAT data.
Initially, the spectral index of the new power-law component was fixed to that obtained for 
\emph{Model E}. The best-fit is acceptable ($\rm{\chi^2_r/dof= 5.94/6}$).
The best-fit absorption value is $\rm{N_H=(2.1_{-1.9}^{+2.3})\times 10^{24}cm^{-2}}$.
If we let the spectral index free, the best-fit values are in agreement with the previously derived values although 
with larger uncertainties: $\rm{\Gamma=2.7\pm0.8}$ and 
$\rm{N_H=(3.9_{-3.6}^{+4.4})\times 10^{24}cm^{-2}}$. 
The power-law component is contributing with a flux of 
F(14-70~keV)$\sf{= (1.5\pm0.4)\times 10^{-11} erg~s^{-1} cm^{-2}}$,
around  $\rm{\sim}$7 times higher than the reflection component. This implies, at 
the distance of NGC\,4102, an intrinsic luminosity of $\rm{L(2-10~keV)=1.4\times 10^{42}erg~s^{-1}}$.
This value is consistent with the \emph{Suzaku}/XHD flux limit. 
Fig. \ref{fig:suz_bat} shows the \emph{Suzaku}/XIS (black and red points below 10 keV) together with
the \emph{Swift}/BAT (green points above 10 keV) data and best-fit models.

\begin{figure}[!t]
\includegraphics[width=1.0\columnwidth]{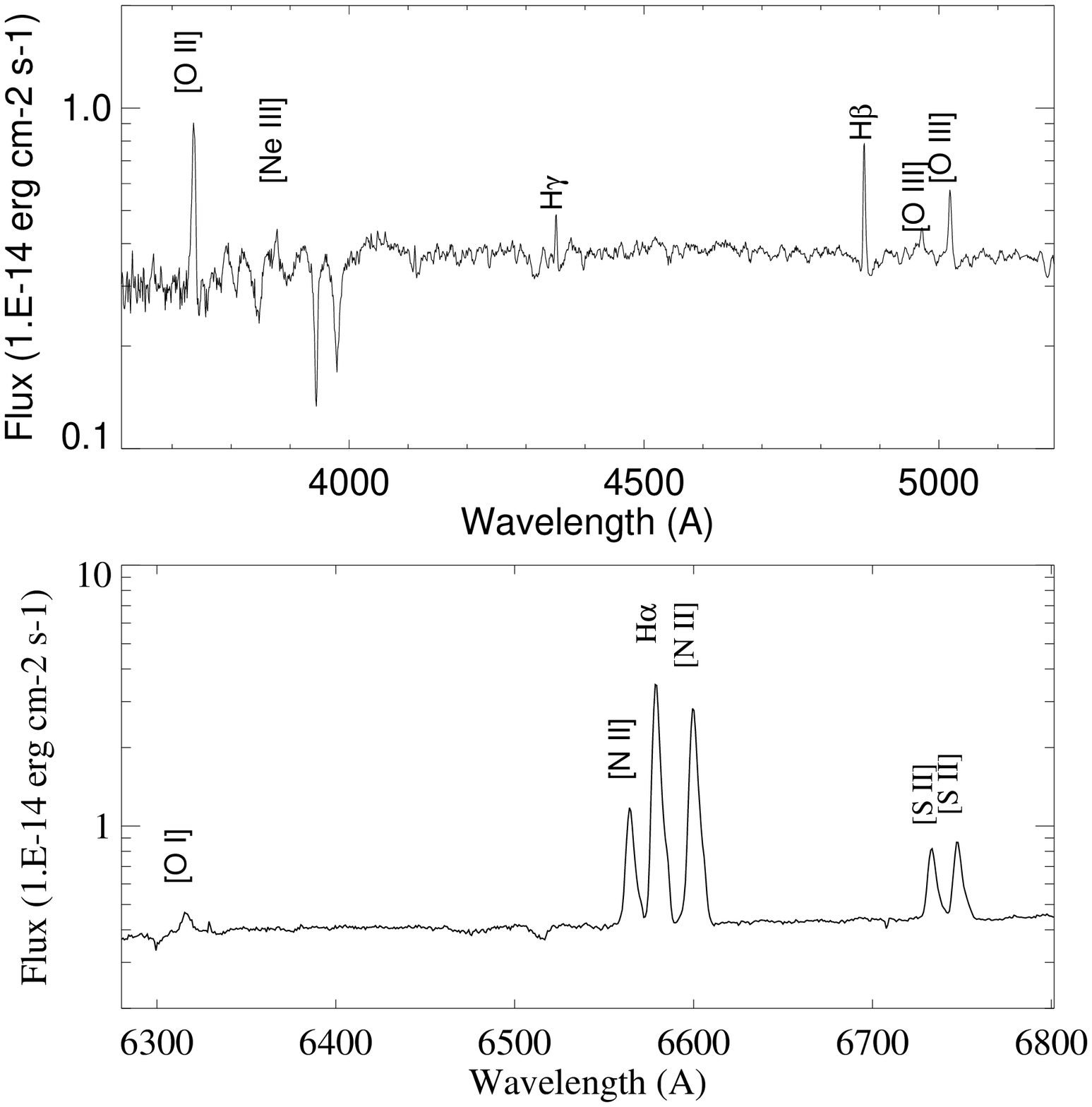}
\includegraphics[height=1.\columnwidth,angle=90]{IR_SPECTRUM.ps}
\caption{Optical spectrum observed with  TWIN in Calar Alto observatory 
(top and middle panels) and near-infrared spectrum observed with LIRIS 
in El Roque de los Muchachos Observatory (bottom panel). The optical and near-infrared spectra  were
taken 1 week apart from each other, and one month after the
\emph{Suzaku} observation. Note that the optical spectra are in arbitrary units since the
absolute calibration could not be done (see text). }
\label{fig:spec-optic}
\end{figure}

\begin{table}[!t]
\begin{center}   
\caption{Optical and near-infrared observed emission lines.
}
\begin{tabular}{lcccc}
\hline \hline
Line 		                &Wavelength		&Flux		       \\ 
                                & ($\rm{\AA}$)      &                      \\ \hline
$\rm{[O~II]~3727\AA}$	&3736.4$\rm{\pm}$0.11	&38.9$\rm{\pm}$1.5    \\
$\rm{[Ne~III]~3869\AA}$	&3876.9$\rm{\pm}$1.01	&4.6$\rm{\pm}$1.8    \\
H$\rm{\gamma}$			&4350.9$\rm{\pm}$0.31	&5.6$\rm{\pm}$0.9     \\
H$\rm{\beta}$			&4873.42$\rm{\pm}$0.07	&19.3$\rm{\pm}$0.8    \\	   
$\rm{[O~III]~5007\AA}$	&5019.07$\rm{\pm}$0.18	&13.1$\rm{\pm}$0.7    \\
$\rm{[O~I]~6300\AA	}$	&6316.34$\rm{\pm}$0.36	&3.6$\rm{\pm}$0.4   \\
$\rm{[N~II]~6548\AA}$	&6564.51$\rm{\pm}$0.01	&39.0$\rm{\pm}$0.3   \\
H$\rm{\alpha}$			&6579.02$\rm{\pm}$0.01  &152.9$\rm{\pm}$0.4   \\
$\rm{[N~II]~6584\AA}$	&6600.09$\rm{\pm}$0.01	&123.9$\rm{\pm}$0.3   \\
$\rm{[S~II]~6716\AA}$	&6733.25$\rm{\pm}$0.06	&21.6$\rm{\pm}$0.3   \\
$\rm{[S~II]~6731\AA}$	&6747.58$\rm{\pm}$0.05	&24.1$\rm{\pm}$0.3   \\ \hline 	   
$\rm{[S~III]~0.907\mu m}$ 	&  9100.5$\rm{\pm}$1.9  &  24.7$\rm{\pm}$4.5  \\
$\rm{[S~III]~0.953\mu m}$ 	&  9562.7$\rm{\pm}$0.5  &  43.4$\rm{\pm}$3.3  \\
He I$\rm{~1.087\mu m}$ 	       	& 10871.3$\rm{\pm}$1.2  &  17.9$\rm{\pm}$2.5  \\
Pa$\rm{\gamma}$			& 10975.5$\rm{\pm}$1.4  &  13.5$\rm{\pm}$3.3  \\
$\rm{[Fe~II]~1.257\mu m}$ 	& 12608.1$\rm{\pm}$0.8  &  14.2$\rm{\pm}$1.7  \\
Pa$\rm{\beta}$ 			& 12859.4$\rm{\pm}$0.9  &  23.2$\rm{\pm}$3.2  \\
\hline 
\end{tabular}
\label{tab:optic-nir}
\end{center}
Observed fluxes expressed in units of $\rm{10^{-15}erg~s^{-1} cm^{-2}}$.
\end{table}

\section{Optical and near-IR spectra of NGC\,4102}

Here we present the optical (TWIN) and near-infrared (LIRIS) spectra contemporaneous
with the X-ray \emph{Suzaku} observation. Note that we are only interested in the 
detection of AGN signatures seen in these spectra while a more detailed study 
is out of the scope of this paper.

\subsection{Optical TWIN spectra}

Optical spectroscopy was obtained in 2009 July 12th using 
the TWIN instrument at the 3.5m telescope in Calar-Alto 
(CAHA\footnote{The Centro Astron{\'o}mico Hispano Alem\'an is operated
jointly by the Max-Planck Institut fur Astronomie and the IAA-CSIC.}, Spain). 
TWIN includes two separate spectroscopic channels (``Blue" and ``Red") behind the common entrance 
slit aperture. Both channels are equipped with a SITe\#22b 14 2048x800 CCD detector having a pixel
size of 15 microns and a plate scale of 0.56 arcsec~pixel$\rm{^{-1}}$.
Two grating settings were used to cover the optical spectral range:  
the gratings T8 and T10 provides a wavelength coverage from
3500 to 5600\AA{} and  5400 to 6700\AA{} with a dispersion of 1.09
and  0.80\AA{}~pixel$^{-1}$, respectively. 
A long slit of 1.2 arcsec width was placed across the galaxy nucleus and oriented 
at $\mathrm{PA}=237^{\circ}$ (close to parallactic angle). The average seeing during the
observations was $\rm{\sim}$1.5~arcsec. Note that the slit width was smaller than the seeing in
order to avoid the degradation of the spectral resolution. 
Two exposures of 1200 s were taken for each grating setting. 
The target  spectra were reduced following standard procedures using the
spectrum reduction package available in IRAF\footnote{IRAF is
distributed by the National Optical Astronomy Observatories, which
are operated by the Association of Universities for Research in
Astronomy (AURA), Inc., under contract with the National Science
Foundation.}. 
For each grating, we built one bias combining five zero time exposures. After overscan and bias subtraction
we divided the image by the normalised flat-field created using a combination of dome flats. Wavelength calibration
was obtained using comparison lamp observations done at the same telescope position as that of the target. 
Flux calibration was performed using two spectrophotometric standard stars (BD+28D42118 and Feige11018). 
Note that the absolute calibration is not accurate enough.

One dimensional spectra, without sky contribution, were extracted from a 3.4 arcsec aperture centred in the
peak of the continuum. They are shown in Fig. ~\ref{fig:spec-optic} (top and middle panels). 
Recombination lines like H$\alpha$, H$\beta$, and H$\gamma$ appear in emission, although 
an absorption component is also detected. Among the forbidden lines, low-ionisation states dominate 
([O I]$\rm{\lambda6300\AA}$, [N II]$\rm{\lambda\lambda6548,~6583\AA}$, 
[S II]$\rm{\lambda\lambda6716,~6731\AA}$, and [O II]$\rm{\lambda3727\AA}$) and 
intermediate ionisation lines like [O III]$\rm{\lambda5007\AA}$ and
[Ne III]$\rm{\lambda3869\AA}$ are also present.
The flux of all lines was measured by fitting a Gaussian
profile to each line (see Tab.~\ref{tab:optic-nir}). In the particular case of
the two [N II] and the H$\alpha$ lines, we de-blended them by modelling the three 
lines simultaneously, constraining our fit by using the same FWHM for all 
of them. In the H$\beta$ and H$\gamma$ cases, we have used two Gaussians to account for
the emission and absorption components of these lines. 
All the lines appear narrow showing widths below 250 km~$\rm{s^{-1}}$ after 
deconvolving the  instrumental profile.
A strong velocity field is seen in the 2-D spectrum at the inner parts of the galaxy, which may produce
wings and assymmetries in the line profiles.

The emission-line ratios used to classify the source are:
\begin{eqnarray}
log([OIII]/H\beta)=-0.17\pm0.05, \\
log([OI]/H\alpha)=-1.63\pm0.06, \\
log([NII]/H\alpha)=-0.091\pm0.005~and \\
log([SII]/H\alpha)=-0.524\pm0.005.  
\end{eqnarray}
These values are consistent with those reported by \citet{Ho97} and \citet{Moustakas06}.
This indicates that the source classification remains unchanged regarding its optical spectrum
when compared to previous observations.
According to the diagnostic diagrams by \citet{Kauffmann03}, NGC\,4102 should be
classified as a LINER source.
We also expect  no line flux variation for those species coming from the NLR, along the timeline between the 
different observations quoted here.

The [O III]$\rm{\lambda5007\AA}$ flux, corrected for internal extinction, can be interpreted as an 
isotropic indicator of the
AGN power of the source \citep{Maiolino98}. In fact, \citet{Ganda06} showed that the [O III] map in NGC\,4102 traces the
outflowing gas of probably the NLR emission in the AGN. They also showed that [O III] emission is 
extended towards the North-West of the nucleus. 
We have compared the fluxes of the optical lines in our spectra with those reported in the literature by \citet{Ho97} and
\citet{Moustakas06}. Some differences are expected due to the use
of different apertures: thus while we used an aperture size of
1.2$\rm{\times}$3.4 arcsec, \citet{Moustakas06} and \citet{Ho97} used an aperture size of 
2.5$\rm{\times}$2.5 arcsec and 2$\rm{\times}$4
arcsec, respectively. Our fluxes are consistent with those reported by \citet{Moustakas06} and
$\sim 30$\% lower than those reported by \citet{Ho97}. Given the flux calibration uncertainties 
mentioned above we have used our spectra for emission line ratios diagnostics and extinction calculations, 
but we have adopted the better calibrated [O III] flux from \citet{Moustakas06}\footnote{They used 
a wide aperture and the slit orientation coincides with the parallactic angle.}.  
The calibrated flux from \citet{Moustakas06} is $\rm{F([O~III])= 1.5 \times10^{-14} erg~cm^{-2}~s^{-1}}$.
We have determined the reddening using the  H$\alpha$/H$\beta$ and H$\beta$/H$\gamma$ flux ratios 
and the extinction curve obtained by \citet{Calzetti94}  to estimate a colour excess of E(B-V)=0.99 \citep[i.e. 
Av$\rm{\sim}$3.05][]{Cardelli89}. This colour
excess implies a de-reddened [O III]$\rm{\lambda5007\AA}$ flux of $\rm{F([O~III])=3.67\times10^{-13} erg~cm^{-2}~s^{-1}}$. 
At the distance of NGC\,4102 it implies a luminosity of $\rm{L([O~III])=1.27\times10^{40} erg~s^{-1}}$.

The X-ray-to-[OIII] ratio, R$\rm{_{X/[O~III]}}$ can be 
used as an indicator of the Compton-thickness of the source. The typical limit of a \emph{Compton-thick}
source  is  R$\rm{_{X/[O~III]}<}$ 0.5 \citep{Maiolino98,Bassani99}.
The  R$\rm{_{X/[O~III]}=0.55_{-0.07}^{+0.06}}$ when using \emph{Suzaku} flux 
measurements, classifying the source as border-line \emph{Compton-thin} source. Using the \emph{Chandra} flux the ratio is 
R$\rm{_{X/[O~III]}= 0.22_{-0.10}^{+0.08}}$, corresponding to a \emph{Compton-thick} source.

\subsection{Near-infrared LIRIS spectrum}

Infrared long slit spectroscopic data in the range 0.9-1.4$\rm{\mu m}$ 
were obtained in 2009 July 3rd with the near-infrared
camera/spectrometer LIRIS \citep{Manchado04} on the 4.2 m William Herschel Telescope. 
The LR\_ZJ (0.9-1.3$\rm{\mu m}$) grism was used.
We followed an ABBA telescope nodding pattern.
The spatial scale is 0.25 arcsec per pixel, and the slit width 0.75 arcsec.
The total observing time was 3000 s obtained from 5 exposures of 600 sec each. 
The infrared spectrum was taken in order to have a view of the central engine 
less affected by obscuration with respect to the optical spectrum. 

The data were reduced following standard procedures for near-infrared 
spectroscopy, using the {\sc lirisdr} dedicated software developed within the IRAF
environment. The basic reduction steps include sky subtraction, flat-fielding,
wavelength calibration and finally the shift-and-add technique to combine
individual frames. The nearby star HIP\,56334 was observed with the same
configuration and similar airmass of the galaxy to perform the telluric 
correction and flux calibration. 

The resulting near-infrared spectrum of NGC\,4102, in the 0.9-1.4 $\rm{\mu m}$ range 
extracted from 1 arcsec (4 pixels) aperture is shown in Fig. \ref{fig:spec-optic} (bottom panel).
Broad permitted lines and coronal lines (e.g. [S IX] and [S VIII]) were not detected.
The spectrum is characterised by intense low ionisation emission lines, consistent
with the LINER optical classification of the source. 
Labels in this figure indicate the securely identified emission lines.  
The spectrum shows both permitted and forbidden lines. Among the latter, the strongest one are 
[S III]$\rm{\lambda\lambda}$ 0.907 and 0.953 $\rm{\mu m}$, although 
other low ionisation lines are also present like 
[S II]$\rm{\lambda\lambda}$ 1.029 and 1.032 $\rm{\mu m}$, 
[Fe II]$\rm{\lambda\lambda}$ 1.257 and 1.321 $\rm{\mu m}$,
and [C II]$\rm{\lambda}$ 0.982 $\rm{\mu m}$. 
Among the permitted lines we detect Pa$\rm{\beta}$ 
and Pa$\rm{\gamma}$. There is also a hint of Pa$\rm{\delta}$. 
We also clearly detect a line at 1.087  $\rm{\mu m}$ which can be 
identified as He I and another line centred at 1.095 $\rm{\mu m}$ 
which can be identified as Fe II. The emission lines fluxes were 
measured using Gaussian fits with an IDL routine
(see Tab. \ref{tab:optic-nir}). 

\section{Discussion}\label{sec:dis}

We have presented the results from the study of 
\emph{Chandra}, \emph{Swift}, and \emph{Suzaku}
data of NGC\,4102 taken in 2003, 2004-2007 and 2009.
The \emph{Suzaku} observations are presented here for the  
first time. We also present the results from optical (TWIN at Calar Alto observatory) 
and near-infrared (LIRIS at El Roque observatory) data 
contemporaneous to the \emph{Suzaku} observations.
The summary of the results so far are:

\begin{itemize}
\item \emph{Suzaku}/XIS spectra of NGC\,4102 are better fitted with a model which consist on:
 (i) a power-law with a spectral index of $\rm{\Gamma=2.3}$, (ii) 
a reflection component, (iii) two Gaussian lines 
at 6.4 and 6.7 keV, respectively, and (iv) a
 thermal component (we used the {\sc apec} model in  {\tt XSPEC}). The broad band spectra
 are also modified by a cold absorber material with $\rm{N_H\sim1\times 10^{21}cm^{-2}}$. 
 This is the so-called \emph{Model E} in Table \ref{tab:fittings} (see also Section \ref{sec:suzakuspec}). 

\item The high FeK$\rm{\alpha}$ equivalent width [EW(FeK$\rm{\alpha)}$=$\rm{680_{-130}^{+110}}$ eV]
strongly supports the \emph{Compton-thick} nature of this source. 
 The FeK$\rm{\alpha}$ emission line is also detected in the \emph{Chandra} spectrum
with the same flux. We also detected the presence of an ionised Fe$\rm{_{XXV}}$ emission line
at 6.7 keV with an EW of $\rm{EW(Fe_{XXV})=160_{-60}^{+60}}$ eV.

\item No emission above 10 keV is detected with \emph{Suzaku}/HXD,
but it is detected with \emph{Swift}/BAT. The latter 
is fitted with a combination of power-law and reflection component (as that 
used in \emph{Suzaku}/XIS data), plus an additional absorption of $\rm{N_H\sim2\times 10^{24}cm^{-2}}$. 
Therefore, the \emph{Swift}/BAT spectrum offers the first direct look to the intrinsic power-law
continuum of this source, which cannot be observed at lower energies because of the large
absorption.

\item The observed flux below 6 keV in the \emph{Suzaku} data is higher than
the \emph{Chandra} flux in the same band.  This variation 
is better described as a decrease of 
the normalisation of the power-law \emph{and} the thermal component by a factor 
of $\rm{\sim}$7. 

\item The optical and near-infrared spectra of NGC\,4102 show signatures 
of a classical Type-2 LINER. The source classification remains consistent 
with previous reported data \citep{Ho97,Moustakas06}. 

\item The object, according to 
the  R$\rm{_{X/[O~III]}}$ ratio is \emph{Compton-thick} during the \emph{Chandra} observation and border-line 
\emph{Compton-thin} during the \emph{Suzaku} observation. 

$\rm{~}$\\
\end{itemize}

In the following sections we discuss the nature of NGC\,4102 according to the
present results.

\subsection{The neutral iron line emission and the Compton-thick 
nature of the source}

The strength of the neutral iron line is quite large, ruling out the
hypothesis that it is originated in transmission dominated material.
Another possibility is that the line is due to reprocessing from cold,
\emph{Compton-thick} matter. Since the observed line is narrow, this material
cannot be associated with the inner region of the accretion disc.
Moreover, the line flux appears to be constant during the \emph{Chandra}
and \emph{Suzaku} observations\footnote{Although we do not have evidence
of variability, it might be possible that the line has changed in
shorter periods of time and by chance it is consistent with being the
same at the time of \emph{Chandra} and \emph{Suzaku} observations.}.
This implies that it has to be produced at a distance larger than 2 pc
from the central source. Using the relation between the X-ray luminosity 
and the size of the BLR derived by \citet{Kaspi05}, we inferred a 
size of the BLR of the source of 1 day-light ($\rm{\sim}$0.001 pc).
This distance implies that the iron line is
formed in the geometrically thick torus rather than the outer regions of
the accretion disc or the BLR.

In this case, the line's high EW rules out inclination angles lower than
60$\rm{^{o}}$, and, assuming half opening angles of the torus of 30-45
$\rm{^o}$, its hydrogen column density must be of the order of $\rm{N_H
\sim 10^{24}cm^{-2}}$ \citep[]{Ghisellini94}. Interestingly, this column
density agrees very well with the one measured with \emph{Swift}/BAT
data ($\rm{N_H\sim 2\times 10^{24}cm^{-2}}$). We therefore believe that,
the strength of the 6.4 keV line, its lack of flux variation, and the
results from the spectral fitting of the \emph{Swift}/BAT data, provide
strong evidence for the presence of a \emph{Compton-thick} torus, at a distance
of $\sim 2$ pc from the central source, which is blocking our view.

\subsection{The ionised Fe$_{XXV}$ emission line}\label{sec:feXXV}

Ionised line emission can be produced by reflection on an ionised  
accretion disc \citep[e.g. MRK\,766,][]{Miller06}. This is particularly
likely for EWs higher than 100 eV. However, if NGC\,4102 is
\emph{Compton-thick} and we have no direct view to the central source at
energies below $\sim 10$ keV, the line emitting region must be located
at a considerable distance from the nucleus to be visible. Another
possibility is that the line is produced by \emph{Compton-thin} gas
which is photoionised by the nucleus. This material  could also be
responsible for the scattered power-law component seen in the
\emph{Suzaku}/XIS energy band. In this case, the EW is expected to be
lower than $\rm{\sim}$ 100 eV \citep{Bianchi09}. Our measurement of the
EW is still consistent, within the error bars, with this possibility. 

\subsection{Absorbers along the line-of-sight}

Apart from the \emph{Compton-thick} absorber at $\rm{\sim}$ 2 pc away from
the central source, the analysis of the \emph{Suzaku}/XIS spectra
suggested the presence of a second, \emph{Compton-thin} absorber with
$\rm{N_H \sim 10^{21} cm^{-2}}$. 

In addition, the optical spectrum analysis yielded a reddening of Av
$\rm{\sim}$3.05. Assuming a Galactic ratio between gas and dust, we
used the relation of \citet{Bohlin78} (i.e., $\rm{N_H=2.2 \times Av
\times10^{21}cm^{-2}}$) to predict a column density of $\rm{N_H\sim 6.7
\times10^{21}cm^{-2}}$, with an uncertainty of $\sim 30$\%. This value
is not consistent with any of the column density estimates of the two X-ray
absorbers. 

Therefore, it is likely that, apart from the torus blocking the view of the central 
source, the Narrow Line Region (NLR) and the soft X-ray emitting region
are also affected by neutral absorbers, albeit of different column
densities. This difference argues against the possibility that these two
regions coincide (see also the discussion in the following section). 
Moreover, they should be located at distances larger than the torus
since it is blocking our line of sight. This argues in favour of this material
being located far away from the source. Most probably, these
two absorbers are associated with the interstellar medium of
the host galaxy. This is in agreement with \emph{HST} imaging
observations of this source\footnote{Images taken from the 
\emph{Hubble Legacy Archive} (HLA), http://hla/stci.edu/hlaview.html.}, 
which reveal the presence of large amount
of dust in the central region of this galaxy \citep[see also][]{Beck10}.


\subsection{The puzzling soft X-ray emission in NGC 4102}\label{sec:soft}

The ``soft excess" seen in the \emph{Suzaku} spectrum is well described
by a thermal model with $\rm{kT\sim0.7~keV}$. This temperature is
consistent with that of thermal component in LINERs \citep{Gonzalez-Martin09A}. This thermal
component is believed to be associated with extended star-forming
regions close to the centre of the host galaxy \citep{Jogee05}. Another possibility is
that the soft X-ray emitting material is located in the Narrow Line
Region (NLR) of the source photonionized by the central source
\citep[e.g. ][]{Bianchi06,Guainazzi07}. \citet{Ghosh08} have used 
\emph{Chandra} data to study the nuclear morphology of NGC\,4102. 
They detected a ``soft" X-rays extended emission, within a radius 
of 3 arcsec, towards the west of the nucleus. They suggested that this 
extended emission could originate in material photoionised by the AGN
at a distance of $\sim 100$ pc. This is consistent with the location of the NLR in 
this object (250 pc), using the relation  between the 
X-ray luminosity and the size of the NLR given by  \citet{Masegosa10}.

However, the variability of the soft-excess flux within $\sim 7$ years
(i.e. between the \emph{Chandra} and \emph{Suzaku} observations) 
suggests that the material responsible for the soft X-ray emission cannot 
be associated with the NLR (the difference in the column density of the 
absorbers ``seen" by this material and the NLR, also argues against this
possibility). Due to its variability, the soft X-ray emitting region must be
located at least within the torus. The material responsible for this
emission may also be responsible for the $\rm{Fe_{XXV}}$ emission line,
and the scattering of the nuclear continuum to the line of sight.
 It is possible that the soft X-ray emission is also nuclear emission, 
scattered to the line of sight, however its temperature is significantly higher 
than the typical temperatures observed in classical AGNs \citep[kT$\sim$0.1 
keV, see e.g.][]{Gierlinski04}. 

In any case it is difficult to understand the origin of the observed
variability in soft X-rays and even more the fact that both the power-law 
continuum and the soft X-ray, ``thermal" component vary with a similar factor 
between the \emph{Chandra} and the \emph{Suzaku} observations. Perhaps
the observed variability could be due to a variable source within the \emph{Suzaku}
extraction region. As already mentioned in Section \ref{sec:obs},
only one source is seen in \emph{Chandra} image within the \emph{Suzaku}
extraction region (CXO\,J120627.3+524303). This source is located at the end of one 
of the spiral arms of the host galaxy.
The total flux of this source is $\rm{F(0.5-10~keV)=4.8\times
10^{-14}erg~s^{-1}cm^{-2}}$, a factor of 40 lower than the power-law plus thermal flux obtained
during \emph{Suzaku} observation ($\rm{F(0.5-10~keV)=1.8\times
10^{-12}erg~s^{-1}cm^{-2}}$). Although it seems to be an extreme variation,  
this has already been observed in the case of the Circinus
Galaxy, when the high flux state of an extremely variable ultra-luminous X-ray source (ULX)
is the most likely explanation for the flux variability of the source \citep{Bianchi02}. 
An X-ray monitoring follow up of the source with high spatial resolution, perhaps 
with \emph{Chandra}, is needed to understand the nature of the ``soft-excess" 
emission and its variability.


\subsection{Why NGC\,4102 is optically classified as a LINER?}

The intrinsic 2--10 keV band luminosity\footnote{This luminosity is obtained 
using the \emph{Swift} spectral fit and it is corrected for 
absorption.} of the source is L(2-10 keV)=1.4$\rm{\times 10^{42}erg~s^{-1}}$. It implies a 
bolometric luminosity of the AGN of  $\rm{L_{bol} \sim 7\times 10^{43} erg~s^{-1}}$, 
assuming a bolometric correction of $\rm{k_{bol}=50}$ \citep[median value derived 
for LLAGN,][]{Eracleous10}. We can also estimate the black hole mass of the source
using the correlation found between the  black-hole mass and the stellar velocity dispersion
by \citet{Tremaine02}. We adopted the stellar velocity dispersion measurements of $\rm{\sigma_{bulge}}$=
174.3 km $\rm{s^{-1}}$ \citep[][]{Ho09b} to estimate a black-hole mass of  
$\rm{M_{BH}=8\times10^{7}M_{\odot}}$. This estimate is consistent with the average value for LINERs 
\citep{Gonzalez-Martin09B}. Finally, using $\rm{\dot{m}_{Edd}=L_{bol}/L_{Edd}=
0.725 (L_{bol}/10^{38})/(M_{BH}/M_{\odot})}$, the derived accretion rate is 
$\rm{\dot{m}_{Edd}=6\times10^{-3}}$. This low accretion rate is also common among LINERs. 

Moreover, the bolometric luminosity, black-hole mass, and accretion rate for this source
are consistent with Type-2 Seyferts \citep[see figures 6 and 7 in][]{Panessa06}. 
However, the optical and near-infrared spectra show low strength of the coronal lines, characteristic 
of LINERs. Therefore, there is still an important open question: why this source 
is optically classified as a LINER and not as a Type-2 Seyfert? 

If we attribute this classification  to the absorption, a special configuration of absorbers, 
partially blocking the continuum emission to reach the NLR has to be claimed. 
We have demonstrated that NGC\,4102 shows a complex structure of absorbers.
However, these complex absorbers are also seen in other Type-2 
Seyferts and, with the current data, we cannot give any evidence supporting this 
special location of the absorbers. The difference found between 
this LINER and Type-2 Seyferts is the steeper spectral index ($\rm{\Gamma=2.3}$),
which is common in other LINERs \citep{Gonzalez-Martin09A}. One possibility is that 
the continuum emission, i.e. the spectral energy distribution (SED) of this source, and 
the accretion mode are different than in other AGN. 

\subsection{The X-ray-to-[O III] ratio as a Compton-thick diagnostic}

Finally, we want to make a remark on the implications of the variable X-ray flux 
detected in this source and the use of  R$\rm{_{X/[O~III]}}$ ratio as an indicator 
of Compton-thickness. The [O~III] comes from the NLR, which is a rather extended
region and, therefore, its emission is constant. The results of this analysis
are consistent with this interpretation, despite of the absolute calibration uncertainties. 
However, this could not be the case for the X-ray flux. The 2--10 keV band flux could 
varies affecting this ratio. For NGC\,4102 the scattering and thermal components 
change in a factor of 7 between the two observations, strongly affecting also the 2--10 
keV flux. These changes can \emph{mimic} a \emph{Compton-thin} source  when the 
source is in a high X-ray flux-state, which it partially does in the case of the \emph{Suzaku} 
observations for NGC\,4102. Therefore, we want to stress that this ratio must be used 
carefully as a \emph{Compton-thick} indicator.  

\section{Conclusions}\label{sec:con}

NGC\,4102 shows indications of \emph{Compton-thick} material associated to the
geometrically thick torus based on the high EW and lack of variability 
of the neutral iron emission line in the \emph{Suzaku}/XIS data, together with the intrinsic 
continuum detected with \emph{Swift}/BAT.  Apart from the \emph{Compton-thick} absorber, 
two more absorbers are derived, from the soft X-rays and the optical spectrum, respectively. 
They are probably associated with material located in the host galaxy. Moreover, we argue that a material 
located within the torus and perpendicular to the plane of the disc could be responsible 
for the observed scattered X-ray continuum component, ``soft-excess", and the ionised Fe$\rm{_{XXV}}$ 
emission lines. To understand the nature of this material an X-ray monitoring follow up of the 
source is required. 

Finally, NGC\,4102 is consistent  with the LINER classification through optical and near-infrared
spectra. However, the accretion rate and bolometric luminosity of this source are consistent 
with other LLAGN like Type-2 Seyferts. We speculate that this classification might be due to a 
different SED due to a different accretion mode, based on the steeper spectral index found.

\begin{acknowledgements}
We thank to the referee for his/her useful comments.
This research has made use of data obtained from the Suzaku satellite, a collaborative mission 
between the space agencies of Japan (JAXA) and the USA (NASA).
This research has made use of data obtained from the Chandra Data Archive, and software provided 
by the Chandra X-ray Center (CXC) in the application package CIAO.
Based on observations collected at the Centro Astron\'omico Hispano Alem\'an (CAHA) at Calar 
Alto, operated jointly by the Max-Planck Institut fur Astronomie and the Instituto de Astrof\'isica de Andaluc\'ia (CSIC).
This article is based on observations made with the  William Herschel Telescope operated in La Palma 
 by the Isaac Newton Group in the Spanish Observatory El Roque de los Muchachos.
OGM thanks M. Guainazzi for useful discussion on this source.
OGM acknowledges support by the EU FP7-REGPOT 206469 and ToK 39965 grants.
I.M and J.M. acknowledge finalcial support from the Spanish grant AYA2007-62190 
and Junta de Andaluc'a TIC-114 and the Excellence Project P08-TIC-03531.
MAM acknowledges the support by the Spanish research project AYA2008-05572.
The Space Research Organization of The Netherlands is supported financially by NWO, 
the Netherlands Organization for Scientific Research.
PE acknowledges financial support from STFC.
\end{acknowledgements}


\begin{thebibliography}{dummy}
\bibitem[Antonucci(1993)]{Antonucci93} Antonucci, R.\ 1993, \araa, 31, 473 
\bibitem[Bassani et al.(1999)]{Bassani99} Bassani, L., Dadina, M., Maiolino, R., Salvati, M., Risaliti, G., della Ceca, R., Matt, G., \& Zamorani, G.\ 1999, \apjs, 121, 473 
\bibitem[Beck et al.(2010)]{Beck10} Beck, S.~C., Lacy, J.~H., \& Turner, J.~L.\ 2010, \apj, 722, 1175 
\bibitem[Bianchi et al.(2002)]{Bianchi02} Bianchi, S., Matt, G., Fiore, F., Fabian, A.~C., Iwasawa, K., \& Nicastro, F.\ 2002, \aap, 396, 793 
\bibitem[Bianchi et al.(2006)]{Bianchi06} Bianchi, S., Guainazzi, M., \& Chiaberge, M.\ 2006, \aap, 448, 499 
\bibitem[Bianchi et al.(2009)]{Bianchi09} Bianchi, S., Guainazzi, M., Matt, G., Fonseca Bonilla, N., \& Ponti, G.\ 2009, \aap, 495, 421 
\bibitem[Bohlin et al.(1978)]{Bohlin78} Bohlin, R.~C., Savage, B.~D., \& Drake, J.~F.\ 1978, \apj, 224, 132 
\bibitem[Brightman \& Nandra(2008)]{Brightman08} Brightman, M., \& Nandra, K.\ 2008, \mnras, 390, 1241 
\bibitem[Calzetti et al.(1994)]{Calzetti94} Calzetti, D., Kinney, A.~L., \& Storchi-Bergmann, T.\ 1994, \apj, 429, 582 
\bibitem[Cardelli et al.(1989)]{Cardelli89} Cardelli, J.~A., Clayton, G.~C., \& Mathis, J.~S.\ 1989, \apj, 345, 245 
\bibitem[Carrillo et al.(1999)]{Carrillo99} Carrillo, R., Masegosa, J., Dultzin-Hacyan, D., \& Ordo{\~n}ez, R.\ 1999, Revista Mexicana de Astronomia y Astrofisica, 35, 187 
\bibitem[Dewangan \& Griffiths(2005)]{Dewangan05} Dewangan, G.~C., \& Griffiths, R.~E.\ 2005, \apjl, 625, L31 
\bibitem[Dickey \& Lockman(1990)]{Dickey90} Dickey, J.~M., \& Lockman, F.~J.\ 1990, \araa, 28, 215 
\bibitem[Dudik et al.(2005)]{Dudik05} Dudik, R.~P., Satyapal, S., Gliozzi, M., \& Sambruna, R.~M.\ 2005, \apj, 620, 113 
\bibitem[Dudik et al.(2009)]{Dudik09} Dudik, R.~P., Satyapal, S., \& Marcu, D.\ 2009, \apj, 691, 1501 
\bibitem[Eracleous et al.(2010)]{Eracleous10} Eracleous, M., Hwang, J. A., \& Flohic, H. M. L.~G.\ 2010, \apj, 711, 796 
\bibitem[Fabbiano(1989)]{Fabbiano89} Fabbiano, G.\ 1989, \araa, 27, 87 
\bibitem[Fukazawa et al.(2009)]{Fukazawa09} Fukazawa, Y., et al.\ 2009, \pasj, 61, 17 
\bibitem[Ganda et al.(2006)]{Ganda06} Ganda, K., Falc{\'o}n-Barroso, J., Peletier, R.~F., Cappellari, M., Emsellem, E., McDermid, R.~M., de Zeeuw, P.~T., \& Carollo, C.~M.\ 2006, \mnras, 367, 46 
\bibitem[Ghisellini et al.(1994)]{Ghisellini94} Ghisellini, G., Haardt, F., \& Matt, G.\ 1994, \mnras, 267, 743 
\bibitem[Ghosh et al.(2008)]{Ghosh08} Ghosh, H., Mathur, S., Fiore, F., \& Ferrarese, L.\ 2008, \apj, 687, 216 
\bibitem[Gierli{\'n}ski \& Done(2004)]{Gierlinski04} Gierli{\'n}ski, M., \& Done, C.\ 2004, \mnras, 349, L7 
\bibitem[Gon{\c c}alves et al.(1999)]{Goncalves99} Gon{\c c}alves, A.~C., V{\'e}ron-Cetty, M.-P., \& V{\'e}ron, P.\ 1999, \aaps, 135, 437 
\bibitem[Gonz{\'a}lez-Mart{\'{\i}}n et al.(2006)]{Gonzalez-Martin06} Gonz{\'a}lez-Mart{\'{\i}}n, O., Masegosa, J., M{\'a}rquez, I., Guerrero, M.~A., \& Dultzin-Hacyan, D.\ 2006, \aap, 460, 45 
\bibitem[Gonz{\'a}lez-Mart{\'{\i}}n et al.(2009a)]{Gonzalez-Martin09A} Gonz{\'a}lez-Mart{\'{\i}}n, O., Masegosa, J., M{\'a}rquez, I., Guainazzi, M., \& Jim{\'e}nez-Bail{\'o}n, E.\ 2009a, \aap, 506, 1107 
\bibitem[Gonz{\'a}lez-Mart{\'{\i}}n et al.(2009b)]{Gonzalez-Martin09B} Gonz{\'a}lez-Mart{\'{\i}}n, O., Masegosa, J., M{\'a}rquez, I., \& Guainazzi, M.\ 2009b, \apj, 704, 1570 
\bibitem[Goulding \& Alexander(2009)]{Goulding09} Goulding, A.~D., \& Alexander, D.~M.\ 2009, \mnras, 398, 1165 
\bibitem[Guainazzi \& Bianchi(2007)]{Guainazzi07} Guainazzi, M., \& Bianchi, S.\ 2007, \mnras, 374, 1290 
\bibitem[Heckman(1980)]{Heckman80} Heckman, T.~M.\ 1980, \aap, 87, 152 
\bibitem[Ho(2009)]{Ho09a} Ho, L.~C.\ 2009, \apj, 699, 626 
\bibitem[Ho et al.(2009)]{Ho09b} Ho, L.~C., Greene, J.~E., Filippenko, A.~V., \& Sargent, W.~L.~W.\ 2009, \apjs, 183, 1 
\bibitem[Ho et al.(1997)]{Ho97} Ho, L.~C., Filippenko, A.~V., \& Sargent, W.~L.~W.\ 1997, \apjs, 112, 315 
\bibitem[Jogee et al.(2005)]{Jogee05} Jogee, S., Scoville, N., \& Kenney, J.~D.~P.\ 2005, \apj, 630, 837 
\bibitem[Kaspi et al.(2005)]{Kaspi05} Kaspi, S., Maoz, D., Netzer, H., Peterson, B.~M., Vestergaard, M., \& Jannuzi, B.~T.\ 2005, \apj, 629, 61 
\bibitem[Kauffmann et al.(2003)]{Kauffmann03} Kauffmann, G., et al.\ 2003, \mnras, 346, 1055
\bibitem[Kinney et al.(1993)]{Kinney93} Kinney, A.~L., Bohlin, R.~C., Calzetti, D., Panagia, N., \& Wyse, R.~F.~G.\ 1993, \apjs, 86, 5 
\bibitem[Magdziarz \& Zdziarski(1995)]{Magdziarz95} Magdziarz, P., \& Zdziarski, A.~A.\ 1995, \mnras, 273, 837 
\bibitem[Maiolino et al.(1998)]{Maiolino98} Maiolino, R., Salvati, M., Bassani, L., Dadina, M., della Ceca, R., Matt, G., Risaliti, G., \& Zamorani, G.\ 1998, \aap, 338, 781 
\bibitem[Manchado et al.(2004)]{Manchado04} Manchado, A., et al.\ 2004, \procspie, 5492, 1094 
\bibitem[Masegosa et al.(2010)]{Masegosa10} Masegosa, J., M{\'a}rquez, I., Ramirez, A., \& Gonz{\'a}lez-Mart{\'{\i}}n, O.\ 2010, arXiv:1011.0865 
\bibitem[Mateos et al.(2005)]{Mateos05} Mateos, S., Barcons, X., Carrera, F.~J., Ceballos, M.~T., Hasinger, G., Lehmann, I., Fabian, A.~C., \& Streblyanska, A.\ 2005, \aap, 444, 79 
\bibitem[Miller et al.(2006)]{Miller06} Miller, L., Turner, T.~J., Reeves, J.~N., George, I.~M., Porquet, D., Nandra, K., \& Dovciak, M.\ 2006, \aap, 453, L13 
\bibitem[Mitsuda et al.(2007)]{Mitsuda07} Mitsuda, K., et al.\ 2007, \pasj, 59, 1 
\bibitem[Moustakas \& Kennicutt(2006)]{Moustakas06} Moustakas, J., \& Kennicutt, R.~C., Jr.\ 2006, \apjs, 164, 81 
\bibitem[Panessa et al.(2006)]{Panessa06} Panessa, F., Bassani, L., Cappi, M., Dadina, M., Barcons, X., Carrera, F.~J., Ho, L.~C., \& Iwasawa, K.\ 2006, \aap, 455, 173 
\bibitem[Panessa \& Bassani(2002)]{Panessa02} Panessa, F., \& Bassani, L.\ 2002, \aap, 394, 435 
\bibitem[Rees(1984)]{Rees84} Rees, M.~J.\ 1984, \araa, 22, 471 
\bibitem[Tremaine et al.(2002)]{Tremaine02} Tremaine, S., et al.\ 2002, \apj, 574, 740 
\bibitem[Tueller et al.(2010)]{Tueller10} Tueller, J., et al.\ 2010, \apjs, 186, 378 
\bibitem[Tully(1988)]{Tully88} Tully, R.~B.\ 1988, Science, 242, 310 
\bibitem[Tzanavaris \& Georgantopoulos(2007)]{Tzanavaris07} Tzanavaris, P., \& Georgantopoulos, I.\ 2007, \aap, 468, 129 
\bibitem[Winter et al.(2009)]{Winter09} Winter, L.~M., Mushotzky, R.~F., Reynolds, C.~S., \& Tueller, J.\ 2009, \apj, 690, 1322 
\bibitem[Winter et al.(2008)]{Winter08} Winter, L.~M., Mushotzky, R.~F., Tueller, J., \& Markwardt, C.\ 2008, \apj, 674, 686 
\end{thebibliography}
\end{document}